\documentclass[useAMS,usenatbib]{mn2e}

\usepackage{graphicx}
\usepackage{times}
\usepackage{natbib}
\usepackage{color}

\newcommand{\apjs}{ApJS}

\newcommand{\apjl}{ApJL}

\newcommand{\aap}{A \& A}

\newcommand{\enzo}{\small{ENZO}}

\begin{document}
 
\title[FRB and magnetic fields] {Probing the origin of extragalactic magnetic fields with Fast Radio Bursts}
\author[F. Vazza, M. Br\"{u}ggen, P. M. Hinz,   et al ]{F. Vazza$^{1,2,4}$ \thanks{E-mail:XX}, M. Br\"{u}ggen$^{2}$, P. M. Hinz$^{2,3}$,   D. Wittor$^{1,2,5}$, N. Locatelli$^{1,4}$, C. Gheller$^{6}$\\
$^{1}$ Dipartimento di Fisica e Astronomia, Universit\'{a} di Bologna, Via Gobetti 93/2, 40121, Bologna, Italy\\
$^{2}$ Hamburger Sternwarte, University of Hamburg, Gojenbergsweg 112, 21029 Hamburg, Germany \\
$^{3}$ Universit\"ats-Sternwarte M\"unchen, Scheinerstr. 1, 81679 M\"unchen, Germany\\
$^{4}$ INAF - Istituto di Radioastronomia, Via Gobetti 101, 40129 Bologna, Italy\\
$^{5}$ INAF - Osservatorio di Astrofisica e Scienza dello Spazio di Bologna, Via Gobetti 93, 40129 Bologna, Italy\\
$^{6}$ CSCS-ETHZ, Via Trevano 131, Lugano, Switzerland.}

\date{Received / Accepted}
\maketitle
\begin{abstract}

The joint analysis of the Dispersion and Faraday Rotation Measure from distant, polarised Fast Radio Bursts may be used to put constraints on the origin and distribution of extragalactic magnetic fields on cosmological scales.  While the combination of Dispersion and Faraday Rotation Measure can in principle give the average magnetic fields along the line-of-sight, in practice this method must be used with care because it strongly depends on the assumed magnetisation model on large cosmological scales. Our simulations show that the observation of Rotation Measures with $\geq 1-10 ~\rm rad/m^2$ in $\sim 10^2-10^3$ Fast Radio Bursts will likely be able to discriminate between extreme scenarios for the origin of cosmic magnetic fields, independent of the exact distribution of sources with redshift. This represent a strong case for incoming (e.g. ALERT, CHIME) and future (e.g. with the Square Kilometer Array) radio polarisation surveys of the sky.

\end{abstract}

\label{firstpage} 
\begin{keywords}
galaxy: clusters, general -- methods: numerical -- intergalactic medium -- large-scale structure of Universe
\end{keywords}


\section{Introduction}
\label{sec:intro}

Fast radio bursts (FRB) are powerful ($\sim \rm Jy$), dispersed and intermittent bursts of radio waves, whose origin has yet to be understood. They are often found at high galactic latitudes and characterized by very large values of dispersion measure (DM), which suggests that they are of extragalactic origin  \citep[][]{2007Sci...318..777L,2013Sci...341...53T,frb_cat}.
According to recent estimates, an FRB may occur at each second within the observable Universe \citep[][]{2017ApJ...846L..27F}. As they traverse the intergalactic medium, the dispersion of radio waves from extragalactic FRB can help detect the missing baryonic matter in the Universe \citep[e.g.][]{2014ApJ...780L..33M}.

\begin{figure*}
\includegraphics[width=0.99\textwidth]{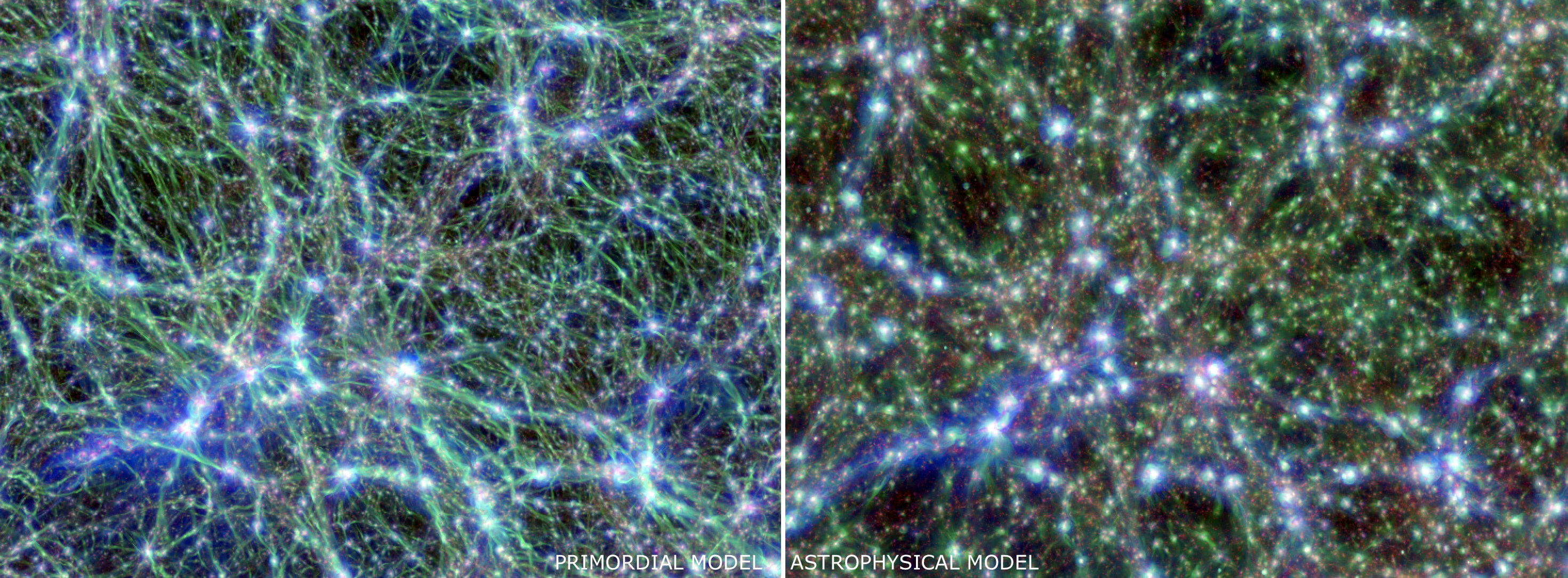}
\caption{3-dimensional renderings of the  projected distribution of dark matter (red), gas temperature (blue) and magnetic field strength (green) at $z=0.02$ for the primordial and astrophysical model investigated in this paper. Each panel is $60 ~\rm Mpc \times 40 ~\rm Mpc $ across, and has a depth of $200 \rm  ~Mpc$ along the line of sight.}
\label{fig:maps}
\end{figure*}

\begin{figure}
\includegraphics[width=0.495\textwidth]{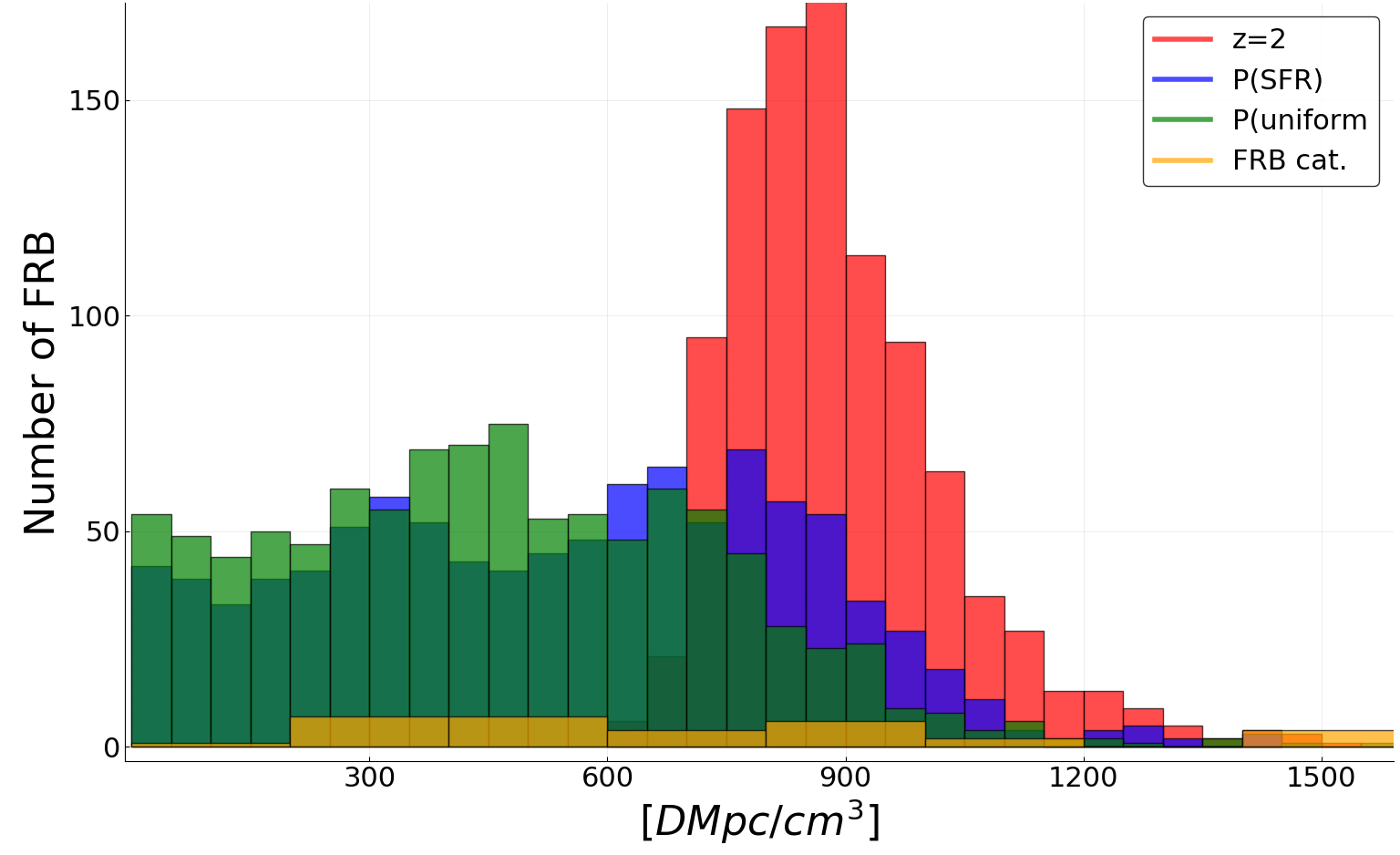}
\caption{Distribution of DM in observed FRB \citep[][]{frb_cat} (after removing from the putative contribution from the Milky Way)  and simulated in this work, by assuming 3 different models for the location of sources (see Sec.\ref{sim}).}
\label{fig:DM_dist}
\end{figure}

If the sources of FRB are at cosmological distances ($z_{\rm FRB} \sim 0.2-2$), their estimated isotropic release of energy is in the range of $\sim 10^{40}-10^{42} \rm \ erg/s$. However, there is no consensus yet concerning their emission mechanisms \citep[e.g.][]{frb1,frb2,frb3}. 
FRB can be used for cosmological parameter studies \citep[e.g.][]{2017arXiv171111277W}, and the simultaneous detection of their Faraday Rotation Measure (RM) and their DM will make it possible to infer the average magnetic field along the line-of-sight \citep[LOS, e.g.][]{2015MNRAS.451.4277D,2017MNRAS.469.4465P}.

Unlike the analysis of RM from polarized radio galaxies, the brightness of FRB allows us to probe lower values of RM and to measure the rotation of their polarization angle with higher accuracy. 
Only a few dozen FRB have been detected so far, and only for a handful of them it has been possible to measure the corresponding RM \citep[][]{frb_cat}. However, the situation is expected to improve dramatically with ongoing \citep[][]{2015MNRAS.454..457P,2017arXiv170604459K} and future \citep[e.g.][ see also ALERT survey with Aperitif]{2013ApJ...776L..16T} radio surveys, specifically designed to detect FRB, with an expected detection rate of a few FRB {\it per day}.

The possible role of FRB to explore extragalactic magnetic fields is particularly important as the  distribution  of  magnetic fields beyond the scale of galaxies and clusters of galaxies is still largely unknown.  In cosmic voids, magnetic fields are constrained to be within upper limits of order $\sim \rm nG$, derived from the Cosmic Microwave Background \citep[][]{PLANCK2015},  and possibly above the lower limits inferred by the absence of Inverse Compton Cascade from distant blazars \citep[e.g.][]{2002ApJ...580L...7D,2015PhRvD..91l3514C}, of order $\sim 10^{-7} \ \rm nG$. In filaments of the cosmic web, only limits at the level of a few $\sim \rm nG$ have been inferred from radio observations \citep[][]{2016A&A...594A..13P,2016PhRvL.116s1302P,vern17}. 
Any detection of magnetic fields beyond galaxies and galaxy clusters will help to explore the origin of cosmic magnetism \citep[][]{va15radio,va17cqg} by distinguishing between primordial processes  in the early Universe \citep[e.g.][]{wi11,2011ApJ...726...78K,sub16}, or more local astrophysical processes related to galaxy formation \citep[][]{donn09,xu09}.

Due to the very weak radio signal expected outside of halos, other methods have been proposed to probe extragalactic magnetic fields, such as by studying Ultra High Energy Cosmic Rays  \citep[e.g.][]{2005JCAP...01..009D,2017arXiv171001353H,2018arXiv180507995B}. Common with this approach is that neither for Ultra High Energy Cosmic Rays nor for FRB the sources are known. In this study, we rely on some very simple assumptions about the distances to the sources and we use new cosmological simulations of extragalactic magnetic fields to study the opportunities and intrinsic limitations that a future large number of detected FRB in Faraday Rotation will offer to the study of the distribution and origin of extragalactic magnetic fields. 
This paper is structured as follows: after describing our simulations and numerical tools in Sec.~\ref{methods}, we present our results in Sec.~\ref{results}. We provide physical and numerical caveats in Sec.~\ref{caveats} before we summarize and conclude our work in Sec.~\ref{conclusion}.

\begin{figure*}
\includegraphics[width=0.99\textwidth]{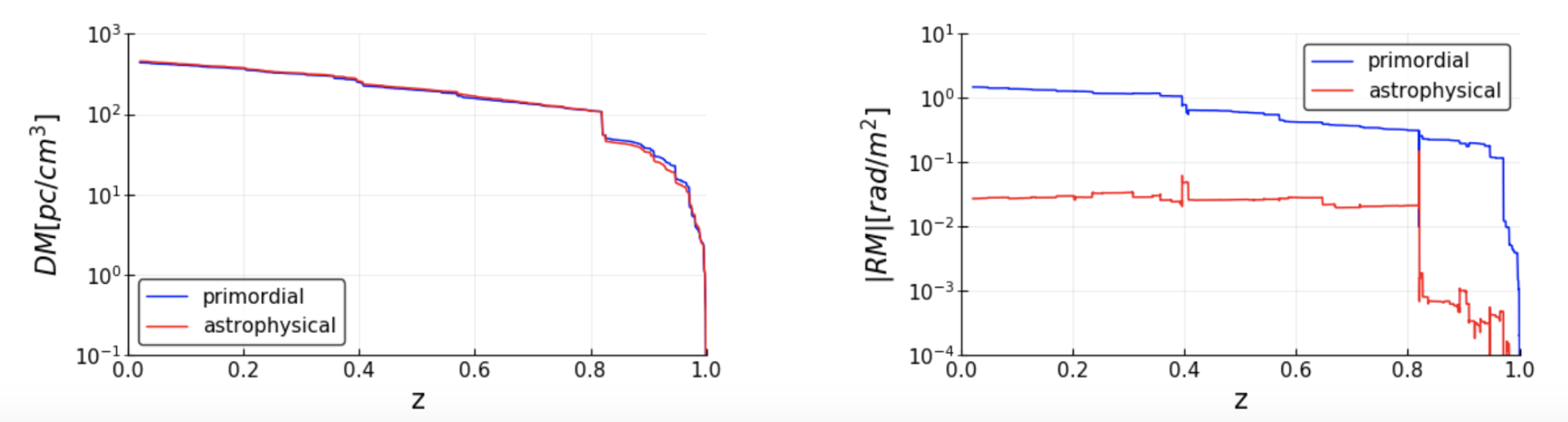}
\caption{DM and RM along one representative LOS in the two models considered in this work, for a sample source located at $z=1$.}
\label{fig:1D}
\end{figure*}

\begin{figure}
\includegraphics[width=0.49\textwidth]{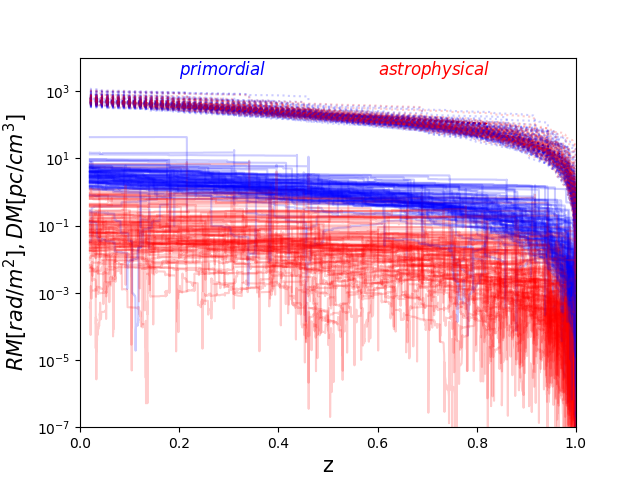}
\caption{Trend of DM and RM for 100 simulated LOS in the two models considered in this work, for sources located at $z=1$. The top dotted lines show the trend of DM, while the lower solid lines give the trend of RM.}
\label{fig:1D_100}
\end{figure}   

\section{Methods}
\label{methods}

\subsection{Numerical simulations}
 \label{sim}
 
We simulated the evolution of dark and baryonic matter as well as cosmic magnetic fields using the {\enzo} code \citep{enzo14}. The simulation covered a $200^3 ~\rm Mpc^3$ (comoving) volume using $2400^3$ cells and dark matter particles, achieving the constant comoving resolution of $83.3 ~\rm kpc/cell$ and the constant mass resolution of $m_{\rm DM}=6.19 \cdot 10^{7} \ M_{\odot}$ per dark matter particle. \\
In order to best bracket uncertainties, we employ here two different models of gas physics and magnetic fields, similar to \citet{va17cqg}.
The first is a non-radiative run with a primordial scenario for magnetic fields, in which we initialised a uniform magnetic field of $10^{-9} ~\rm G$ (comoving) already at the start of the simulation ($z_{\rm in}=38$) and let it evolve under the effect of gravity and magneto-hydrodynamical forces. In the second model, the
astrophysical scenario, we use radiative simulations including a simple prescription for feedback from active galactic nuclei (AGN). Now the initial seed field is 100 times smaller ($10^{-11} ~\rm G$ comoving) and the gas can cool via equilibrium cooling assuming a primordial composition. When cooling pushes the gas density above $n_{\rm th}=10^{-2} \ \rm part/cm^3$ (typical of cool core galaxy clusters), bipolar thermal jets with a fixed budget of $E_{\rm AGN}=10^{58}$ erg of thermal energy (and $E_{\rm B}=1\% \ E_{\rm AGN}$ of magnetic energy) are launched to mimic the large-scale effects of AGN self-regulation of cluster atmospheres. Even though the second scenario also includes a weak primordial seed field, for simplicity we refer to it as "astrophysical" as the resulting  present-day magnetic fields in large-scale structures are dominated by the injection and amplification of magnetic fields seeded at low redshifts by AGN.\\

In previous work, we have shown how this implementation is fairly effective in reproducing the $M-T$ scaling relation on the high-mass end and on the low-mass end of observed clusters, large-scale profile of density and temperature in clusters \citep[][]{va13feedback,scienzo16}, as well as the observed distribution of radio relic sources \citep[][]{va15radio}. 
For a discussion on the comparison of this approach to more sophisticated models based on supermassive black hole particles, we refer the reader to \citet{va17cqg}. 

Fig.~\ref{fig:maps} shows the large-scale distribution of dark matter, gas temperature and magnetic fields in the two scenarios.  The differences between the thermal and magnetic properties of the halos are small across runs, while the differences (especially in magnetic fields) become increasingly large moving into the more rarefied environment of filaments and voids. We find that in the primordial case underdense regions show a $\sim 10-10^2$ higher magnetic field values than the run with feedback. 

In all runs, we assumed a $\Lambda$CDM cosmological model, with density parameters $\Omega_{\rm BM} = 0.0478$, $\Omega_{\rm DM} =
0.2602$,  $\Omega_{\Lambda} = 0.692$, and a Hubble constant $H_0 = 67.8$ km/sec/Mpc \citep[][]{2016A&A...594A..13P}. 

 In Sec.~4.4 we will also discuss further simulations that include other models for seed fields.

\begin{figure*}
\includegraphics[width=0.45\textwidth]{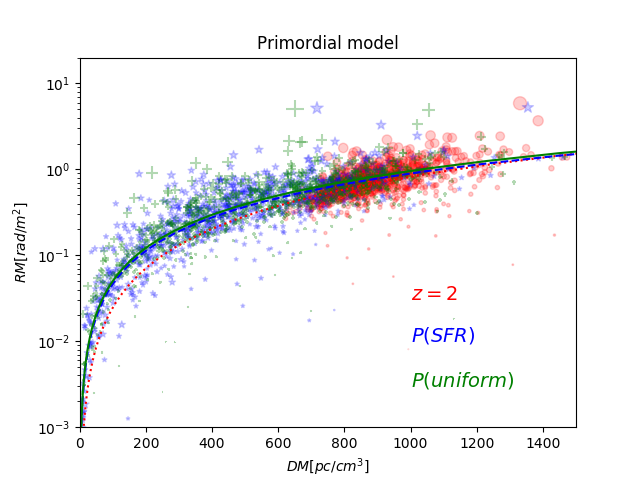}
\includegraphics[width=0.45\textwidth]{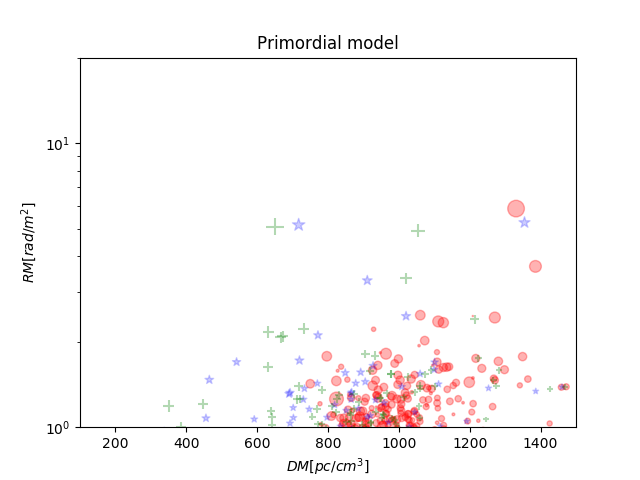}
\includegraphics[width=0.45\textwidth]{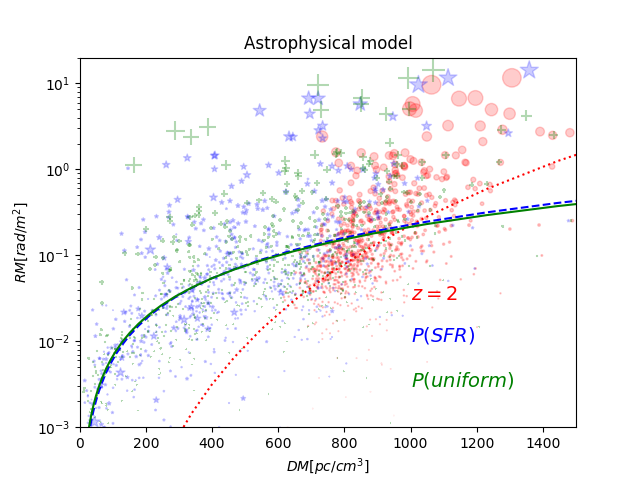}
\includegraphics[width=0.45\textwidth]{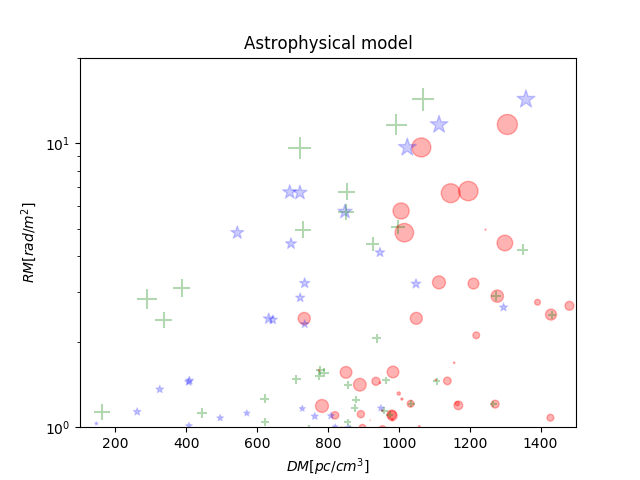}
\caption{RM vs DM for 1000 LOS in the primordial scenario (top panels) and in the  astrophysical scenario (bottom panels) scenario. The different colors and symbols denote different models for the distribution of FRB. The right plots give the close-up view of the high $\rm |RM|$ region of the plots, which might be observable with SKA-MID. The size of markers linearly scales with the average strength of the magnetic field along each LOS.}
\label{fig:dmrm}
\end{figure*}

\subsection{Simulated Rotation and Dispersion measures}
\label{dm_dist}
We compared the RM and DM of both resimulations by extracting 1000 one-dimensional beams of cells along the $z$-axis of the simulation at random $x,y$-positions (identical for the two runs). 
Since the redshift distribution of observed FRB is still unknown
\citep[e.g.][]{2016MNRAS.461..984O,2017ApJ...846L..27F,2018arXiv180401548W}, we tested three different procedures to locate FRB sources in the simulation. 
In a first, simple model, we assumed all FRB are located at redshift $z=2$ ($d_L=15.8$ Gpc). In a second model, we randomly extracted 
the redshift of sources from a distribution that follows the observed cosmic star formation rate (SFR) as a function of redshift, which is well approximated by $P(SFR) \propto z^{3}$ for $z \leq 2$ \citep[e.g.][]{2014ARA&A..52..415M}, assuming a maximum distance of $z=2$. In a third model, we have randomly drawn sources assuming they are uniformly distributed in comoving radial distance. 
In order to generate LOS from
$z=0.02$ to $z=2.0$, we concatenated 26 beams of $2400$ cells each taken from outputs at different redshift, for a total comoving distance of $\approx 5.2$ Gpc. Given that our computational box has a fixed comoving volume, our corresponding sampling $\Delta z$ is discontinuous, and goes from $\Delta z \approx 0.045$ at $z=0.02$ to $\Delta z \approx 0.2$ towards $z=2$.  
In order to avoid artefacts caused by the periodic repetition of structures along each LOS from $z=2$ to $z=0.02$, we concatenated 26 different LOS crossing the $200^3 ~\rm Mpc^3$ box from random positions at increasing redshift.
Of course, the discrete sampling in redshift causes artifacts (especially in the magnetic field distribution) at the boundaries of each box. However, considering that the fraction of cells close to boundaries is $8 \cdot 10^{-4}$ of the total number of cells for each LOS, the artifacts introduced at the edges have a negligible effect on the resulting DM and RM.

Each of the 1000 LOS consists of 62,400 cells with a continuously interpolated
distribution of redshift, and physical values of lengths, density and magnetic fields based on this redshift distribution. 

The DM for each LOS is defined as:
\begin{equation}
DM  \ \rm [kpc/cm^3]= \int_0^{d_{FRB}} \frac{n_e(z)} {(1+z) \ cm^3} \frac{dl}{\rm kpc} ,
\label{eq:dm}
\end{equation}
where $d_{\rm FRB}$ is the assumed comoving distance of each FRB, $z$ is the redshift of each cell and $n_e$ is the physical electron density of cells, assuming a primordial chemical composition ($\mu=0.59$) of gas matter everywhere in the volume. 
The values of DM along a simulated LOS are given in the first panel of Fig.~\ref{fig:1D}: the differences between the same LOS in our two models are extremely small as the gas density distribution on such scales is hardly affected by cooling or feedback. 

We should point out that the standard analysis of DM
\citep[e.g.][]{2003ApJ...598L..79I} often assumes a uniform and completely ionised distribution of baryons, which allows to  simplify Eq.~\ref{eq:dm} to $z_{FRB} = \rm DM/1200 ~pc/cm^3$ (which is valid up to $z=2$ with a very small $\sim 2\%$ scatter). However, from the 
distribution of DM from our reference $z=2$ model in Fig.~\ref{fig:DM_dist} we can clearly see that
even when all FRB are located at the same distance, the observed scatter in DM is large ($\sim 30\%$) as most of the cosmic volume is underdense, and the number of overdense structures that are crossed by each LOS introduces a substantial deviation on the final DM of each source. 

The RM for each LOS is defined as:

\begin{equation}
  RM \ \rm[rad/m^2]=812 \int_0^{d_{FRB}} \frac{B_{\rm ||}}{\rm \mu G} \cdot \frac{n_e}{\rm (1+z)\ cm^3} \frac{dl}{kpc},
\end{equation}
where $\rm ||$ denotes the component of the magnetic field parallel to the LOS. The observed DM and RM of a single extragalactic FRB has contributions from the local environment of the source, its host galaxy, the intergalactic medium and finally the Milky Way \citep[e.g.][]{2014ApJ...790..123A}. In what follows, we only consider the contribution from the intergalactic medium.\\

The panels in Fig.~\ref{fig:1D} shows the integrated DM and $\rm |RM|$ from $z=1$ to $z=0.02$ for the same LOS. 
While the DM is consistent at all redshifts within a few percent in the two cases, the $\rm |RM|$ clearly differs, with the marked tendency of the primordial model to show a higher $\rm |RM|$ at all redshift. Only close to the sites where magnetic fields are injected by AGN, localized in the highest density peaks of the simulated volume, the $\rm |RM|$ in the two models become similar. 
This trend is consistently found in most LOS: the RM difference between the same set of LOS is typically larger than the cosmic variance within models. Fig.~\ref{fig:1D_100} shows the sample variance for 100 LOS integrated from $z=0$ to $z=1$, where a clear segregation in RM (unlike in DM) is measured in the two scenarios. 
This behavior indicates already that any inversion of the $\rm |RM|$/DM relation to obtain the average magnetic field along the LOS will give different results, depending on the physical model for the origin of extragalactic magnetic fields. \\

Fig.~\ref{fig:DM_dist} gives the distribution of the DM for our three source models for FRBs, compared with the presently known distribution of FRBs based on the {\it FRBcat Catalog}(http://frbcat.org), where we considered the DM of each FRB after removing the putative $\rm DM_{\rm MW} $ from the Milky Way \citep[][]{frb_cat}.  The latter contribution is derived using the electron density model by \citet{2002astro.ph..7156C}, while the intrinsic contribution from the hosts of FRB is unknown \citep[see e.g.][for a recent discussion]{2018arXiv180401548W}, and is therefore neglected, consistently with our simulated data. 
While none of our models are successful in reproducing the observed distribution of DM in real FRB, which shows no clear dependence on DM, the model based on the cosmic SFR yields a slightly larger fraction of $\rm DM \geq 800-1000$, similar to observed statistics \citep[][]{frb_cat}.

\subsubsection{Higher resolution data}

While these simulations are among the largest and, certainly in the underdense regions, most resolved cosmological simulations of extragalactic magnetic fields, our computational grid is too coarse to properly model the growth of magnetic fields in galaxy clusters, which is governed by a small-scale dynamo \citep[e.g.][]{bm16}. The simulation of a small-scale dynamo requires a fairly large effective Reynolds number, which is still difficult to achieve, certainly in simulations of cosmological scales \citep[][]{va18mhd}. 
In order to include the role of magnetic field amplification within massive halos for a subset of our fiducial runs, we replaced the magnetic field values from cells located in overdensities typical for clusters by those from higher resolution runs ($\sim 4 \rm ~kpc/cell$) \citep[][]{va18mhd}. This experiment is only performed for the non-radiative simulation with a primordial magnetic field. 

In particular,  we identified all cells with a gas density larger than the typical gas density at the virial radius of self-gravitating halos  ($\rho \geq 50 \rho_{c,b}$ where $\rho_{c,b}$ is the cosmic baryon critical density at a given redshift) and randomly replaced both their density and magnetic fields by randomly drawing from LOS taken from the adaptive mesh refinement simulation. This nested procedure approximates a full AMR simulation with $\sim 4-5$ levels of refinement on top of our baseline $83.3 ~\rm kpc$ resolution.

\section{Results}\label{results}

We start by analysing the distribution of observable DM and $\rm |RM|$, as shown in Fig.~\ref{fig:dmrm}. 
In  Table~\ref{tab0}, we give  the best-fit  parameters for a simple $\rm log_{\rm 10}\left(|RM|\right)=\alpha + \beta \cdot \rm log_{\rm 10}(DM) $ relation, for all models and source distributions considered here. 
 
The two models of magnetic fields fill the ($\rm |RM|$,DM) plane in different ways. The primordial model tends to produce larger RM values for $\rm DM \leq 600~\rm pc/cm^3$ while the astrophysical model produces a larger scatter of values at any DM.
Remarkably, in the primordial scenario the different distribution of sources yields the same best-fit relation, of the kind

\begin{equation}
\frac{|\rm RM|}{\rm rad/m^2} \approx 10^{-4} \left(\frac{\rm DM}{\rm pc ~cm^3}\right)^{1.3 \div 1.5} ,
\end{equation}
with exact parameters given in Table~\ref{tab0}. A very similar relation, albeit with a larger scatter, also holds for the astrophysical models, with the exception of the $z=2$ source model, which shows a steeper slope ($\beta \sim 4.6$). 
This can be understood because $z=2$ is close to the peak of simulated AGN activity. Hence LOS are more likely to cross magnetised bubbles released by AGN which consequently yield larger RM.\\

The situation is reversed if we only look at $\rm |RM| \geq 0.1 ~rad/m^2$, which is the only part of the RM space that observations in the foreseeable future might be able to probe. As expected, the highest RMs are found in the astrophysical model. This is a result of higher densities that are found in radiative simulations as well as a result of stronger magnetic fields close to higher density regions that experience AGN feedback. Also we find a much larger scatter of RMs in the astrophysical model. 

If we go to low RMs ($\rm |RM| \geq 1-10 ~rad/m^2$) which might enter the parameter space probed by SKA-MID, the  distribution of FRB in the ($\rm |RM|$,DM) plane will have the potential to discriminate among such extreme scenarios: while in the primordial case we expect a small scatter around $\sim 1-2 ~\rm rad/m^2$ for sources with $\rm DM \geq 1000 ~pc/cm^3$, we expect possible variations up to $\sim 2$ orders of magnitude in RM ($\rm |RM| \sim 0.2-20 ~\rm rad/m^2$), for the same range of DM, {\it irrespective of the exact distribution of sources}. In particular, while in the primordial scenarios we expect that $\rm |RM|$ values from the cosmic web should be clustered around $DM \approx 900 ~\rm pc/cm^3$ and $\rm |RM| \approx 1 ~rad/m^2$, factor $\sim 5-10$ times larger RMs may be expected in the astrophysical scenarios, nearly independent of their DMs. 

Already with a few tens of sources in this ($\rm |RM|$,DM) range, it will be possible to falsify  models of extragalactic fields, even without a detailed knowledge of the intrinsic redshift of the sources.

\begin{figure}
\includegraphics[width=0.45\textwidth]{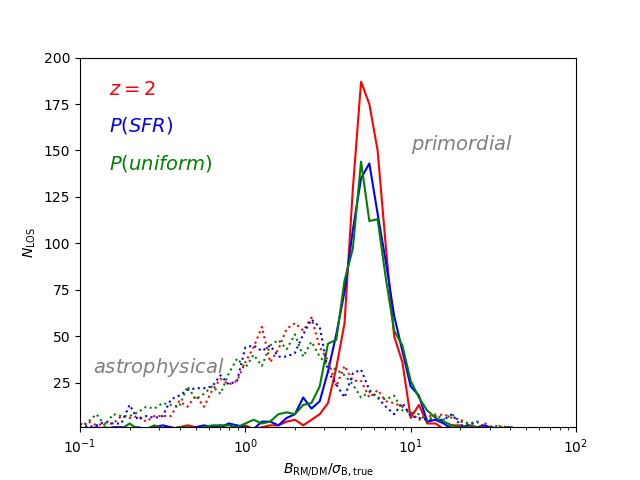}
\caption{Distribution of the ratio between the average magnetic field inferred from the B(RM/DM) relation and the true one for our simulated LOS. }
\label{fig:btrue/blos}
\end{figure}

Our results point towards the possibility of studying the origin of cosmic magnetism using FRB through the observed scatter of RMs. However, they also imply that FRB are not a trustworthy probe of the magnetisation of the intergalactic medium along the LOS.
Even without taking into account the potential role of the unknown contribution of DM and RM by their host galaxies, the 
 ($\rm |RM|$,DM) relation given by our simulations, irrespective of the 
 exact distribution of sources, is affected by a large scatter. 
 We combined the mock  RM and DM values of all simulated FRB in our sample, and estimate the magnetic field along each LOS as:
 
\begin{equation}
B_{\rm RM/DM} = \frac{1000 ~\rm pc/cm^3 }{812~\rm rad/m^2} \cdot \frac{\rm |RM|}{\rm DM}.
\label{eq:bdmrm}
\end{equation}
This can be compared to the real rms value of magnetic fields along each LOS ($\sigma_{\rm B,true}$).
the results are shown in Fig.~\ref{fig:btrue/blos}: in both scenarios and fairly independently on the redshift distribution of sources, the {\it intrinsic} scatter in $B_{\rm RM/DM}/\sigma_{\rm B,true}$ is large, $\sim 1$ order of magnitude in the primordial case and $\sim 2$ in the astrophysical scenario. Moreover, in the primordial case the peak of the observed distribution of $\sigma_{\rm B,true}$ is systematically biased high by a factor $\sim 4-5$ with respect to the real value measured along each LOS. This bias is nearly absent ($\leq 2$) in the astrophysical scenario, yet the very large scatter in the reconstructed values makes it too uncertain to rely on Eq.~\ref{eq:bdmrm}. \\

We note that \citet{2016ApJ...824..105A} have suggested to recalibrate Eq.~\ref{eq:bdmrm} in order to recover the average magnetic field along LOS probed by FRB, by introducing a weighting factor, $\langle 1+z \rangle f_{\rm DM}$, where $f_{\rm DM}$ is the fraction of DM produced by the gas phase in the $10^5 ~\rm K \leq T \leq 10^7 ~\rm K$ range, which was mostly responsible for the observed $(\rm|RM|,DM)$  values found in their simulated LOS. 
In particular, they discussed one specific model where extragalactic magnetic fields are predicted based on the phenomenological use of magnetic field amplification efficiency factors, derived from turbulence-in-a-box simulations \citep[][]{ry08}, and are not the result of full MHD simulations.\\ 

The differences in our models can account for the reported differences in the use of the ($\rm |RM|,DM$) relation to measure cosmic magnetism: in the \citet{2016ApJ...824..105A} model the magnetisation of filaments (assumed in post-processing as a regular function of cosmic environment) dominates the signal along the LOS. Hence, by assessing the typical relation between the cosmic density of filaments and their magnetisation, it is possible to calibrate the ($\rm |RM|,DM$) relation to infer magnetic fields with good precision. Conversely, our results show that there is no single correction factor to calibrate the ($\rm |RM|,DM$) relation. Firstly, even in the primordial scenario we observe that the local dynamics of filaments can introduce a significant scatter in their final magnetisation, even within the same range of overdensities. Secondly, in the assumed  primordial scenario, the contribution to the RM along LOS up to z=2 is not entirely due to gas in the $10^5-10^7 ~\rm K$ range as in \citet{2016ApJ...824..105A}. Hence, there is no single gas phase that can be used to reliably calibrate the ($\rm |RM|,DM$) relation. Finally, AGN activity can introduce an even larger source of scatter in the observable ($\rm |RM|,DM$) relation because, for a given cosmic overdensity, the local magnetic field can vary, based on the preceding AGN activity.

\begin{table}
\label{tab0}
\caption{Best fit parameters for the $\rm log_{\rm 10}|RM|=\alpha + \beta \cdot \rm log_{\rm 10}DM $ relations  (with $1 \sigma$ deviation) in the various models and runs explored in this work. }
\centering \tabcolsep 5pt 
\begin{tabular}{c|c|c|c}
B-field model & source model & $\alpha$ & $\beta$ \\  \hline
Primordial & $z=2$ & $-4.65 \pm 0.46 $& $1.52 \pm 0.15$ \\
Primordial & P(SFR) & $-3.90 \pm 0.08$ & $1.28 \pm 0.03$ \\
Primordial & P(uniform) & $-3.90 \pm 0.09$ & $1.28 \pm 0.03$ \\
Astrophysical & $z=2$ & $-14.62 \pm 1.14$ & $4.66 \pm 0.38$\\
Astrophysical & P(SFR) &  $-5.35 \pm 0.17$ & $1.57 \pm 0.07$\\
Astrophysical & P(uniform) & $-5.22 \pm 0.19$ & $1.50 \pm 0.07$\\
\end{tabular}
\label{tab0}
\end{table}

\begin{figure}
\includegraphics[width=0.45\textwidth]{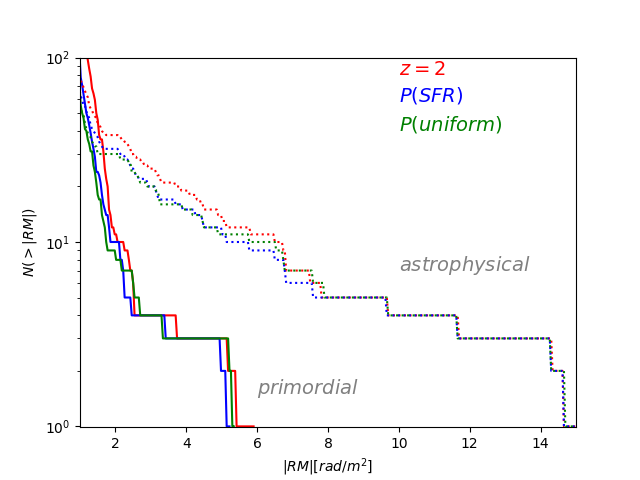}
\caption{Distribution of $\rm |RM|$ for $\geq 1~\rm rad/m^2$ data in our sample.}
\label{fig:rm_dist}
\end{figure}

\begin{figure}
\includegraphics[width=0.495\textwidth]{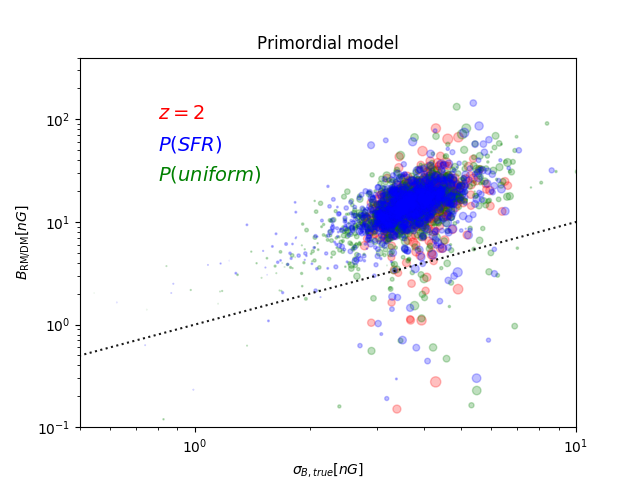}
\includegraphics[width=0.495\textwidth]{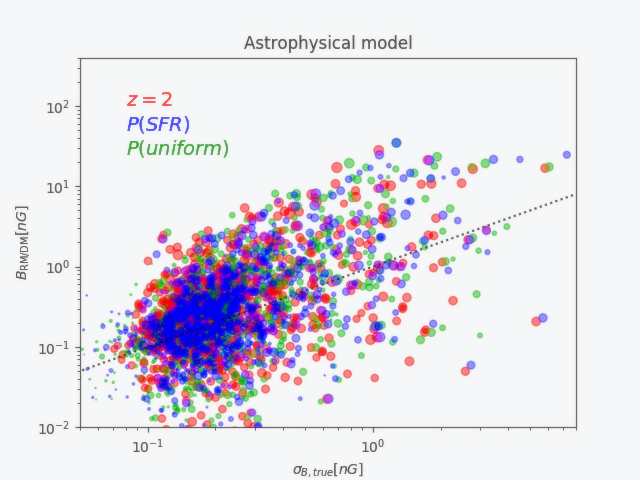}
\caption{Relation between the rms magnetic field along the LOS and the magnetic field which is inferred from the combination of RM and DM, comparing primordial (top) and astrophysical (bottom) scenarios for the different investigated source models. The size of dots correlates with the DM along each LOS.}
\label{fig:frb_distr}
\end{figure}

\section{Physical and numerical caveats}\label{caveats}

\subsection{Astrophysical uncertainties}
An unavoidable limitation of our analysis is that the contributions to DM and RM by the FRB hosts  remain unknown, and they are expected to strongly depend on the morphology of the host galaxy as well as on the exact location of the FRB within it \citep[e.g.][]{2018arXiv180401548W}.
While the Milky Way halo is expected to contribute to  ${\rm DM}_{\rm MW}\sim 40 \pm 15 ~\rm pc/cm^3$ \citep[e.g.][]{2015MNRAS.451.4277D}, hosts of FRB are expected to contribute more to the DM \citep[e.g.][]{2015RAA....15.1629X}. 
For example, a host DM contribution of $\sim 250 ~\rm pc/cm^{3}$ has been suggested for the
repeating FRB 12110, which is localized in the star-forming region of a $z = 0.19$ dwarf galaxy \citep[][]{2017Natur.541...58C}. 
For sources located within $z \leq 2$, \citet{2018arXiv180401548W} suggested that a DM in the range of  ${\rm DM}_{\rm host} \sim 50-500 ~\rm pc/cm^3$ might be expected for reasonable variations of possible progenitors and location within host galaxies. \\

Another unavoidable source of uncertainty in radio observations results from the  Galactic foreground. Studying the redshift dependence of RM in a large sample of quasars,  \citet{2017ARA&A..55..111H} recently suggested that up to  $\sim 10^4-10^5$ measurements of RMs may be necessary to tell apart Galactic from extragalactic contributions to the dispersion of RMs in distant objects. FRBs may ease this requirement because they may be distributed more evenly at low redshift ($z \leq 1$) than quasars. Moreover, it is expected that our knowledge of the three-dimensional structure of the Galactic magnetic field will  improve at the same pace as RM statistics, thus improving templates of the Galactic foreground screen. For example, the combination of  several set of observables (such as extragalactic RMs, PLANCK polarisation data, galactic synchrotron emission and observed distribution of ultra-high energy cosmic rays) through Bayesian inference  already allows considerably improvements in the reconstruction of Galactic magnetic fields across a wide range of scales
\citep[e.g.][for a recent review]{2018arXiv180502496B}.
    
\subsection{Numerical uncertainties}    
Because of the finite numerical resolution, any simulation will be agnostic about fluctuations below a given cell size. In the case of MHD simulations, an additional problem is that small-scale magnetic field fluctuations are seeded by the {\it inverse} cascading of magnetic energy from small to large scales, if the flow is in the small-scale dynamo regime. This is observed in the case of galaxy clusters, simulated at sufficiently high ($\leq 10 ~\rm kpc$) spatial resolution \citep[e.g.][]{va18mhd}. The constant resolution of $83.3  ~\rm kpc/cell$ (comoving) is sufficient to represent voids, sheets and filaments, where previous studies found no evidence of small-scale dynamo amplification driven by structure formation even at higher resolution \citep[][]{va14mhd}. \\

\begin{figure}
\includegraphics[width=0.495\textwidth]{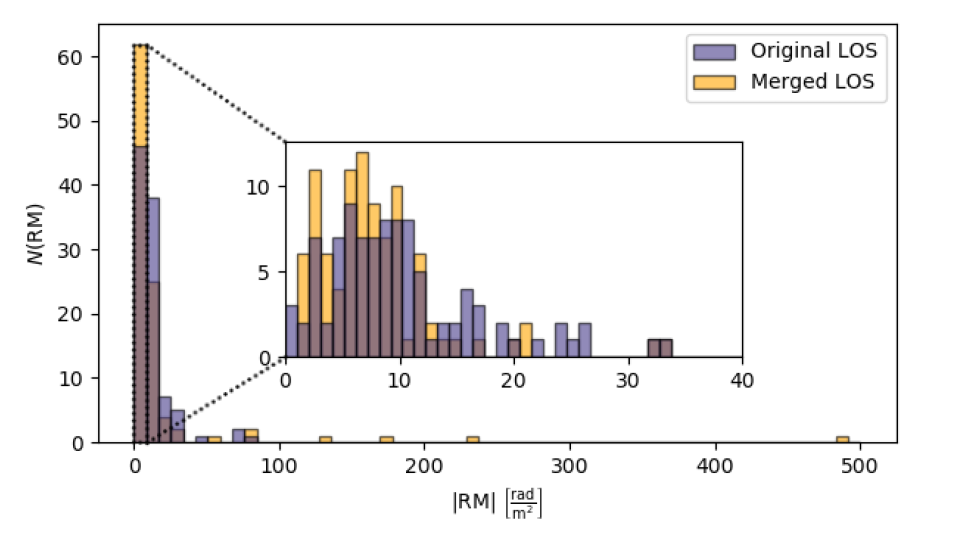}
\caption{Distribution of the RM for 100 simulated LOS, where limited to high-density regions we replaced the magnetic field values of the simulation used in the main paper (with a fixed spatial resolution of $83$ kpc comoving) with highest resolution data from a AMR run simulation of a $10^{15} M_{\odot}$ galaxy cluster (with the peak resolution of $3.9$ kpc comoving). The RM obtained using this combined dataset ("merged LOS") are contrasted with the original RM in the simulation. } 
\label{fig:clust}
\end{figure}

As discussed in Sec. 2.2.1, we have investigated this issue in more detail by replacing the magnetic field and gas density values in 100 of our simulated LOS in the primordial scenario with much higher resolution data ($4 ~\rm kpc$) taken from our recent adaptive-mesh refinement simulations \citep[][]{va18mhd}. 
The resulting distribution of RM is shown in Fig.~\ref{fig:clust}: the volume fraction occupied by cluster-like regions is small in our random distribution of LOS, and beside a few ${\rm |RM|} \sim 250-500 \rm ~rad/m^2$ values (found in LOS clearly piercing the centre of galaxy clusters) the overall statistics of RM values is unchanged (similar results are found for the DM statistics). 
This is further confirmed by computing the fractional contribution to DM and ${\rm |RM|}$ 
from cells with a gas density higher than the virial one along our LOS: cells in the cluster-like environment  are found to contribute on average only to a $\sim 2.5-5 \ \%$ of the observed DM, and to $\sim 15.6 \ \%$ of the observed ${\rm |RM|}$. \newline

The low contamination of galaxy clusters to the overall distribution of DM and RM values for long LOS is consistent with the expected number of galaxy clusters from cosmology. In recent work, \citet{2018arXiv180407063Z} estimated the counts of 
galaxy clusters with masses $\geq 10^{13} \ \rm ~M_{\odot}$ that are expected for $\rm deg^2$ as function of redshift. Their integrated counts from $z=1$ to $z=0$ yields a total of $\approx 493 ~\rm clusters/\mathrm{deg}^2$. Normalized to $\sim 0.003 ~\rm deg^2$ area covered by our simulated beams, this gives an average of $n_{\rm cl} \approx 1.48$ clusters per beam. Assuming that each cluster crossing takes on average $l_c=\rm 1 ~Mpc$, we can estimate an overall $DM_{\rm CL} = 10^2 \cdot n_{\rm cl}\cdot l_{\rm cl} \cdot \langle n \rangle $ contribution to the total DM from $z=1$ sources, while the contribution from the uniform smooth gas from $z=0$ to $z=1$ is $DM = 3.3 \cdot 10^{3}  \rm Mpc \cdot \langle n \rangle$. Hence, we can estimate an average contribution of $DM_{\rm cl}/DM \sim 4.4 \ \%$ from clusters, which is in the range of what we measure in our data ($\sim 2.5-5 \ \%$).

We conclude that, in general, galaxy clusters do not have a significant effect on the distribution of DM and $|\rm RM|$.\\

\subsection{Finite resolution of radio observations}

A further, important limitation to consider in our modeling is that we assume an {\it infinitely small} 
beam size, i.e. we consider for each simulated FRB the effect of a
one dimensional beam of DM and RM values, while in reality an observation with a {\it finite} beam size will see the convolution of a distribution of values along the integration path from the FRB to the observer. This is an unavoidable assumption of any finite volume method, and in this particular case we are assuming that the structure of magnetic fields and density fluctuations below our cell resolution are negligible compared to the resolved scales. This is reasonable, as the typical Jeans scale for density fluctuations in the intergalactic medium is 

\begin{equation}
\lambda_J \approx 1.08 \ \rm Mpc \cdot \sqrt \frac{T/(10^4 \ \rm K)}{\Omega_m (1+\delta)(1+z)} ,
\end{equation}
and the simulated power spectrum of density 
fluctuations show very little structure below $\leq 200 ~\rm kpc$ even at higher resolution \citep[e.g.][]{2015MNRAS.446.3697L}. Moreover, also the magnetic field power spectrum in the entire cosmic volume shows little structure below $\leq 1 ~\rm Mpc$ \citep[][]{va17cqg}.

\begin{figure}
\includegraphics[width=0.495\textwidth]{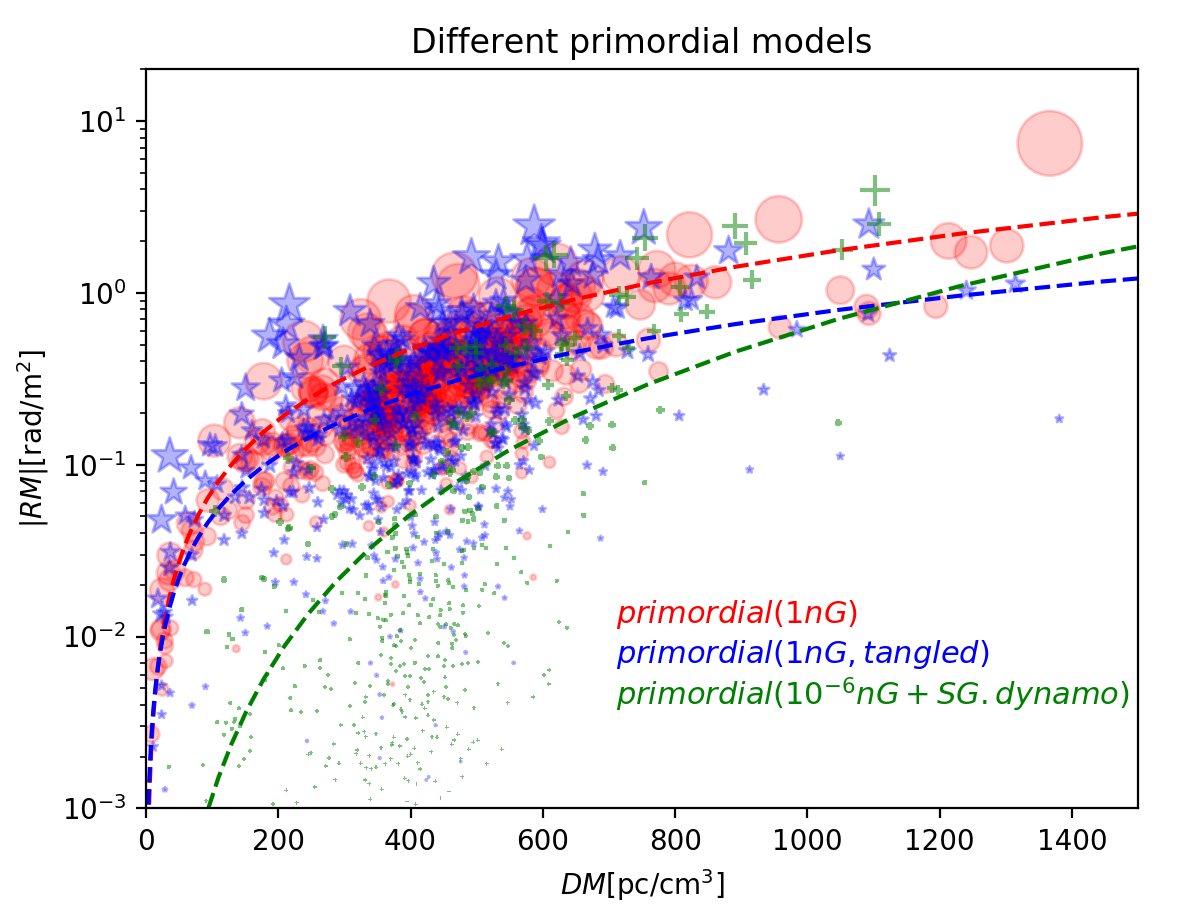}
\caption{RM vs DM for 500 LOS in three more variations of the primordial scenario (see Sec. 4.4 for details).  The different colors and symbols denote different models for the distribution of FRB. The size of markers linearly scales with the average strength of the magnetic field along each LOS. } 
\label{fig:more}
\end{figure}   

\subsection{Additional primordial magnetic models}

The results shown so far were obtained for two extreme scenarios for the origin of extragalactic magnetic fields. In \citet{va17cqg} we presented 20 model variations (for a smaller $85 ~\rm Mpc^3$ volume with identical spatial/mass resolution and cosmology)  of extragalactic magnetic fields, and here we present and additional tests derived from this suite of simulations. In particular, we compared:  a) the identical non-radiative primordial model as in the rest of the paper; b) a primordial model in which we impose a initially  tangled distribution of magnetic fields, using the Zeldovich approach outlined in the Appendix of \citet[][]{va17cqg}. In this approach we initialise magnetic fields that correlate with initial density perturbations by generating magnetic field vectors perpendicular to the 3-D gas velocity field of the Zeldovich approximation (this way ensuring the $\nabla \cdot \vec{B} \equiv 0$ by construction). 
The resulting initial magnetic fields have the same power spectrum of the initial velocity fluctuations and is normalized such that $\sqrt{|<B^2>|} = B_0=1 ~\rm nG$  
c) A  primordial model starting from the much lower seed field value of $B_0=10^{-16} ~\rm G$ (comoving) but featuring a sub-grid model for a small-scale dynamo \citep[see Appendix of][]{va17cqg}. In this model we convert a small fraction $\eta_B$ ($\leq 5-10\%$)of the kinetic energy of the gas solenoidal velocity  into magnetic fields. This approach is based on  \citet{ry08} and the fitting formulas given by \citet{fed14} that allow us to estimate the effective value of $\eta_B$ as a function of the flow Mach number. Moreover, we ensure that the newly generated magnetic field is parallel to the direction of the gas vorticity.  This model is motivated by the attempt of including a scenario in which small-scale vorticity is generated within cosmic filaments, on scales which presently cannot be directly resolved by cosmological MHD simulatons (see. \citealt{va14mhd} for a discussion).

This way we produced analogous sets of 500 LOS for each model.

The resulting ($\rm |RM|,DM$) relation is shown in Fig.~\ref{fig:more}.  We find that the initial field topology has a small effect on the best-fit relation for ($\rm |RM|,DM$), and mainly increases the scatter by a little . The RMs and DMs from FRBs at $z \leq 2$ is relatively independent of the exact topology of the initial magnetic field, while it does depend on the field normalisation.   Conversely, a scenario in which the primordial seed field is very low and the magnetisation is the result of efficient small-scale dynamo amplification even on the scale of filaments \citep[e.g.][]{ry08}, results in a steeper  ($\rm |RM|,DM$) relation that falls in between the primordial and the astrophysical model. We conclude that the observed distribution of $|RM|$ is mostly determined by the intermediate overdensity structures of the cosmic web (e.g. filaments), whose magnetisation would be systematically impacted by the presence of high primordial fields.
By linearly rescaling the level of $|RM|$ observed in our data for lower seed fields, we conclude that  virtually no detection of Faraday Rotation from FRBs should be possible for seed fields below $B_0 \leq 0.1 ~\rm nG$, even if their initial topology is tangled.
If the seed fields are much weaker, a systematic detection of Faraday Rotation from the most dispersed ($\rm DM \geq 10^3 ~\rm pc/cm^3$) FRBs should be possible only if some level of small-scale dynamo amplification occurs on $\ll 50 ~\rm kpc$ scales. Dynamo amplification on these scales are presently hard to directly simulate with cosmological simulations. No systematic detection of Faraday Rotation from extragalactic FRBs, or a large scatter in the observed ($\rm |RM|,DM$) relation for large samples would suggest an astrophysical origin for the magnetisation of large-scale structures.

\section{Observational perspectives and conclusions}\label{conclusion}

We have investigated whether Fast Radio Bursts can be used to study the magnetisation of the Universe \citep[][]{2016ApJ...824..105A}, complementary to the study of radio synchrotron emission from the cosmic web \citep[e.g.][]{va15radio,vern17}, Ultra-High Energy Cosmic Rays \citep[e.g.][]{2003PhRvD..68d3002S,hack16} or blazar halos \citep[e.g.][]{2002ApJ...580L...7D,2009ApJ...703.1078D}. \\

Our analysis suggests that the use of the RM-DM relation to infer the magnetisation of the intervening matter is not straightforward: the RM-DM relation is typically affected by a large scatter, e.g. at $\rm DM=10^3 \rm ~pc/cm^3$ we typically find a scatter of a factor  $\sim 10$ for the RM in the primordial scenario, and of $\sim 10^2$ for the astrophysical case. 
Moreover, due to the non-Gaussian statistics of magnetic field fluctuations in long LOS \citep[e.g.][]{2017PhRvL.119j1101M}, we find that a simple inversion of the RM-DM relation systematically overestimates  the real rms value of magnetic fields along the LOS by a factor $\sim 5-7$ in the primordial case, and by a factor of $\sim 2$ in the astrophysical scenario. In both cases, the scatter is large ($\sim 10-10^2$). \\
Consequently, inferring the average magnetisation of the intergalactic medium from the combination of observed RM and DM of a few FRB will be prone to large errors, making this procedure quite unreliable. However, for large ($\sim 10^2-10^3$) samples and DM in FRBs there is hope to discriminate at least between a purely primordial and a purely astrophysical scenario, as the dispersion of values in the primordial case is predicted to be $\sim 10 \ \%$ of the dispersion of values in the astrophysical scenario. 
Remarkably, discriminating between models based on the dispersion of the detected RM values 
should be possible independent of the distance distribution of sources.

While the DM is only mildly affected by gas physics,
the RM depends strongly on the assumed magnetic field model, and has a wider distribution of values in astrophysical scenarios due to the intermittent nature of AGN feedback.  We conclude that no systematic detection of Faraday Rotation from extragalactic FRBs, or a large scatter in the observed ($\rm |RM|,DM$) relation for large samples of FRBs will imply an astrophysical origin for magnetic fields on large scales.\\

Research on the use of FRB for all kinds of cosmic tomography is still in its infancy, as several  experiments are gearing up to detect large numbers of FRB with DM and RM, such as the Canadian Hydrogen Intensity Mapping Experiment (CHIME) which will observe from 400 to 800 MHz, simultaneously using 1024 beams \citep[][]{2017arXiv171102104N}, or the Aperture Tile in Focus (APERITIF) experiment on the Westerbork telescope, which will observe at 1.4 GHz using $\sim 10^3$ beams \citep[][]{2014htu..conf...79V}. 
When fully operational, also the Square Kilometer Array is expected to produce a breakthrough in the study of FRB: even if extrapolations into its sensitivity range are non trivial, realistic estimates suggest that up to 
one FRB per second over the entire sky will be observable using the SKA-MID2 \citep[][]{2017ApJ...846L..27F}.
Therefore, when numbers of detected FRB will increase to a few thousands per year, their statistical analysis will make it possible to use FRB as a probe of extragalactic magnetic fields and of the origin of cosmic magnetism.

\section*{Acknowledgements}
The cosmological simulations described in this work were performed using the {\enzo} code (http://enzo-project.org), which is the product of a collaborative effort of scientists at many universities and national laboratories. We gratefully acknowledge the {\enzo} development group for providing extremely helpful and well-maintained on-line documentation and tutorials. Most of the analysis done in this work was performed with the Julia code (https://julialang.org).\\
F.V. and D.W. acknowledge financial support from the ERC Starting Grant "MAGCOW", no.714196.  We acknowledge the  usage of computational resources on the Piz-Daint supercluster at CSCS-ETHZ (Lugano, Switzerland) under project s701 and s805. We thank F. Zandanel, J. L. Han and T. Akahori or useful scientific feedback, and our anonymous referee for valuable suggestions.

\bibliographystyle{mnras}
\bibliography{franco,franco2}

\end{document}